\RequirePackage{fix-cm}
\documentclass[twocolumn]{svjour3}          
\smartqed  
\usepackage{graphicx}
\usepackage{caption}
\usepackage{url}
\begin{document}

\title{Prediction error-driven memory consolidation for continual learning. On the case of adaptive greenhouse models
\thanks{GS has received funding from the European Union’s Horizon 2020 research and innovation programme under the Marie Sklodowska-Curie grant agreement No. 838861 (Predictive Robots). Predictive Robots is an associated project of the Priority Programme "The Active Self" of the Deutsche Forschungsgemeinschaft (DFG, German Research Foundation). GS has also received funding from the EU-H2020 Open Cloud Research Environment (OCRE) project and the Marie Curie Alumni Association as part of the European Open Science Cloud initiative, for running this project on GPU accelerated computing resources on the cloud. GS and LM have received funding from the DFG project on initiation of international collaboration "Adaptive Architectures for Transferability of Greenhouse Models". LM has received funding from the Project PROSIBOR. The project "Development of a sensor based intelligent greenhouse management system (PROSIBOR (FKZ: 2815701315)" is supported by funds of the Federal Ministry of Food and Agriculture (BMEL) based on a decision of the Parliament of the Federal Republic of Germany via the Federal Office for Agriculture and Food (BLE) under the innovation support programme.\textit{Author contribution}: GS and LM have designed and implemented the computational models, the experiments and part of the data pre-processing. LM and US have collected and pre-processed the data from the greenhouse facilities.}
}


\author{Guido Schillaci         \and
        Luis Miranda \and Uwe Schmidt  
}


\institute{Guido Schillaci
\at
        The BioRobotics Institute\\
        and Department of Excellence in Robotics \& AI\\ 
        Scuola Superiore Sant'Anna\\
        Pisa, Italy \\
         \email{guido.schillaci@santannapisa.it}
           \and
           Luis Miranda and Uwe Schmidt \at
           Department of Biosystems Engineering  \\ Humboldt-Universit\"{a}t zu Berlin\\ Germany
}

\date{Received: date / Accepted: date}

\maketitle

\begin{abstract}
This work presents an adaptive architecture that performs online learning and faces catastrophic forgetting issues by means of episodic memories and prediction-error driven memory consolidation. In line with evidences from the cognitive science and neuroscience, memories are retained depending on their congruence with the prior knowledge stored in the system. This is estimated in terms of prediction error resulting from a generative model. 

Moreover, this AI system is transferred onto an innovative application in the horticulture industry: the learning and transfer of greenhouse models. This work presents a model trained on data recorded from research facilities and transferred to a production greenhouse.

\end{abstract}

\section{Introduction}

Adaptivity is about adjusting behaviours or beliefs to achieve novel objectives or to respond to unexpected circumstances.
Of crucial importance for biological systems, adaptivity is one of the most challenging capabilities to implement in artificial systems. 
In order to address this quest, developmental robotics takes inspiration from models of human development and from principles of brain functioning. \cite{asada2009cognitive,krichmar2019neurobiologically}. 
Indeed, infant brains are continuously exposed to rich and novel sensorimotor experience while morphological and environmental conditions are changing. 
As an example, a motor skill acquired at a certain point in time -- e.g. sitting up, manipulating toys -- needs to be re-adapted as the proportion of growing body parts change and as other capabilities emerge. 

The scientific community converges on considering the somatosensory cortex of the human brain as playing a role in the implementation of adaptive body representations \cite{holmes2004body}. These representations are formed along the rich sensorimotor information the individual is exposed to, while interacting with its surroundings.
Evidences suggest that experienced sensorimotor contingencies and action-effect regularities are stored in the brain, allowing later processes of anticipation of sensorimotor activity. This has been shown to be crucial for adaptive behaviours,
perception \cite{friston2012prediction}, motor control \cite{adams2013predictions}, memory \cite{ergo2020reward,fountas2020predictive} and many other cognitive functions \cite{hohwy2013predictive,schillaci2016}, and has inspired a wide range of computational models for artificial systems
\cite{cangelosi2015developmental,eppe2019,schillaci2020intrinsic}.
However, despite the promising results in robotics and AI, a number of challenges still remain open. Among these, there is the question about how adaptivity can be leveraged in \textit{lifelong learning} systems. Although there is an increasing understanding of how biological systems balance the integration of new knowledge while retaining past experience, an implementation of such strategies in artificial systems is still arduous.

In mammals, memory is composed of multiple systems supported by different structures in the brain \cite{squire2015memory}. One of these systems, i.e. episodic memory, is crucial for adaptive behaviours, as well as for other cognitive functions such as planning, decision-making and imagination \cite{murty2016episodic}. Memory traces are stabilised in the brain after their initial acquisition through memory consolidation \cite{Traversa2013}. 
Consolidation occurs at different levels in the brain, including a faster, synaptic (hippocampal) level and a slower, more stable (neocortical) system level. System consolidation seems to be driven by the hippocampus, which reorganises its stored temporal and labile memories into more stable traces in the neocortex \cite{squire2015memory}. The rate of consolidation seems to be influenced also by the congruence between prior knowledge and the information that is going to be stored \cite{van2012schema}. Recent studies suggest that if the information to be learned is consistent with prior knowledge, neocortical consolidation can be more rapid \cite{mcclelland2013incorporating,squire2015memory}.
In other words, the way memory is updated seems to be dependent on the extent new information is likely to be formed  \cite{exton2015updating,simon2017brain,sinclair2018surprise}. Moreover, consolidated memories are not static imprints of past experiences, but are rather malleable and can be updated or reconsolidated \cite{roscow2019behavioural,sinclair2019prediction,jang2019positive}.
A key component of this process seems to be the capability of the brain to evaluate a prediction error, or a surprise signal, which would be necessary for destabilising and reconsolidating memories. Evidences suggest also that formation and consolidation of long-term memories are supported by sleep, where experienced events are likely to be reactivated \cite{born2012system}.
The rate of memory consolidation is also clearly dependent on the developmental stage of the individual, as infants show weaker retention of experience compared to adults, reflecting a tendency of the brain to save new learning at that age \cite{gomez2017infants}. Among the aforementioned factors, stress and other emotional conditions can impact episodic memories and memory consolidation \cite{shields2017effects}.

The present work brings a twofold contribution to this special issue. Firstly, it advances the state-of-the-art on continual learning in artificial systems. In particular, it proposes an online learning framework implementing an episodic memory system, in which memories are retained according to their congruence with the prior knowledge stored in the system. This congruence is estimated in terms of prediction error resulting from a generative model.

Secondly, it demonstrates that some of the paradigms of developmental robotics and of brain-inspired computational modelling can be transferred from laboratories to innovative applications. In particular, we apply this research in an application for the horticulture: the transfer of climate models, which are designed to protect the plants and increase the crop yield, from research to production greenhouse facilities.

\subsection{AI Transfer: adaptive greenhouse models}
Continual learning, i.e. the capability of a learning system to continually acquire, refine and transfer knowledge and skills throughout its lifespan, has represented a long standing challenge in machine learning and neural network research \cite{parisi2019continual}. 
Training neural networks in an online and prolonged fashion without caution typically rises \textit{catastrophic forgetting} issues \cite{mcclelland1995there}. Catastrophic forgetting describes the overwriting of previously learned knowledge that occurs when a model is being updated with new information. Researchers have been trying to tackle this issue through different strategies \cite{chen2018lifelong,shin2017continual,kirkpatrick2017overcoming}.
Approaches to prevent catastrophic forgetting include consolidating past knowledge initially present in a short-term memory system into a long-term memory one \cite{mermillod2013stability}, or employing an episodic memory system \cite{mcclelland1995there} that maintains a subset of previously experienced training samples and replays them, along with the new samples, to the networks during the training. 
This paper adopts a mixed approach which uses episodic memory replay and prediction-error driven  consolidation to tackle online learning in deep recurrent neural networks. 
Importantly, this work aims at transferring this AI strategy onto an application for the innovative greenhouses industry.

Greenhouses are complex systems comprising technical and biological elements. Similarly to robots, their state can be measured and modified through control actions, for instance on the internal climate conditions. Modelling the mappings between different sensors and between control actions and resulting measurements allows to anticipate the effects of an intervention upon the greenhouse conditions, to better plan further control actions and, ultimately, to increase crop yield. Several studies can be found in the horticulture literature showing that neural networks can model different processes occurring in a greenhouses, including internal climate \cite{fitz2011neural} and
yield \cite{ehret2011neural,salazar2014tomato}.

Experiments by \cite{miranda2019b} used multilayer perceptrons to predict time series in greenhouses, particularly leaf tissue temperature, transpiration and photosynthesis rates of a tomato canopy. The authors used chained simulations to generate predictions of several time steps, using 3 time steps for all input signals. A more thorough investigation of the time steps needed to predict time series inside a greenhouse is given by \cite{miranda2018}, who points that a static selection of time steps gives poor results after three historical steps (15 minutes) in the inputs. This is due to different time constants in the physical system, with several inputs showing very fast variations and others being dumped. These experiments suggest that more elaborated models are needed to account for the different historical time steps needed to make a prediction: varying for each input and for times of day and seasons.

However, adaptive models have received little attention (e.g. \cite{speetjens2009towards}), despite their potential impact in several applications in the field. 
Indeed, the possibility to adapt can facilitate the transfer of models from research facilities to the production greenhouses. In a preliminary study \cite{miranda2019}, we showed that a learning architecture characterised by deep recurrent neural networks and an episodic memory system can enable the portability of greenhouse models. The model exposed to a big amount of data recorded from a research greenhouse can be transferred to a production facility, requiring less amount of data from the new greenhouse setup. This approach can have a high impact on the greenhouse industry, as it would allow to design and train optimal models at research greenhouses and quickly re-adapt them to different production facilities and crops.

Here, we extend our previous study \cite{miranda2019} by introducing a more efficient memory consolidation strategy and by analysis different aspects of the architecture. As in the previous work, we train a computational model for estimating the transpiration and photosynthesis of a hydroponic tomato crop by using measurements of the climate. The models are trained and tested using data from two greenhouses in Berlin, Germany. Thereafter, the adaptive model is fed with data from a production greenhouse in southern Germany, near Stuttgart, where other tomato varieties were grown under different irrigation and climate strategies.

\section{Methodology}
The computational model adopted here consists of a deep neural network, in part composed by Long Short-Term Memory (LSTM) layers, characterised by two outputs -- transpiration and photosynthesis -- and a time series (of fixed in length) of six sensor values as inputs. In particular, climate data (air temperature, relative humidity, solar radiation, CO2 concentration) and temperature of two leaves are used as sensor data. The model is used to predict transpiration and photosynthesis rates from the sequence of sensor data. Anticipating these information allows better control of the climate and, consequently, an increase of the yield. This part is however not covered by this study.

The samples have been pre-recorded from three different greenhouses (hereon, GH1, GH2 and GH3), with a rate of one multi-sensors measurement every 5 minutes. GH1 and GH2 are research greenhouses located in Berlin. Recordings have been carried out during several years: 2011 to 2014 for GH1, and 2015 to 2016 for GH2.
GH3 is a production greenhouse located near Stuttgart, Germany. Data from 2018 was obtained for this greenhouse.




We test two models\footnote{Diagrams of the structure of the models, as well as the source code, can be found here: \url{https://github.com/guidoschillaci/online_lstm_episodic_memory}.}. In both models the inputs consist of \textit{fixed-length time series} of six sensors data (air temperature, relative humidity, solar radiation, CO2 concentration, temperature on leaf 1, temperature on leaf 2). In particular, the first model (M1) takes as input a window of 288 subsequent samples from the six sensors, corresponding to one full day of recordings, given that samples are captured every 5 minutes. The second model (M2) takes as input a window of 576 subsequent 6D samples, corresponding to two full days of recordings\footnote{One day included a full daylight-night shift in the data. We avoided dealing with smaller portion of time windows, and chose instead multiples of one day. Considering the good performance of the system using two days input data, , as described in the result section, we did not increase much the input size to allow a clearer differentiation between memory consolidation strategies.}. Output consists of a 2D vector representing the transpiration and photosynthesis rates recorded at the same instant when the time window ends.

Datasets are 
prepared so that input-output training samples can be sequentially extracted, to simulate an online learning process. For both models, the first training phase includes 5 cultivation years (2011 to 2014) from GH1. Subsequent phases include cultivation year 2015 (GH2), 2016 (GH2) and lastly the commercial greenhouse GH3 regarding cultivation year 2018. In all cases the time series are truncated during the winter production pauses.

For model M1, this results in 26197 training samples from GH1 (2011 to 2014) exposed sequentially to the learning process. After all samples are covered, the model is exposed to 7079 samples from GH2, (2015) and to 5566 samples from GH2 (2016). Finally, the model is exposed to 1153 samples from GH3 (2018).
During each of these training phases, performance of the learning system is estimated by computing the mean squared error (MSE) on test datasets extracted from the corresponding greenhouse. In particular, test datasets consist of 1377 samples (1/20th of the GH1 training dataset size) for GH1, 372 samples for GH2 (2015), 292 samples for GH2 (206) and finally 60 samples for GH3.

In another experiment, model M2 is trained and tested on smaller datasets, defined by wider input windows  (two days, or 576 samples). In particular, M2 is exposed, in sequence, to 24949 training samples (tested on 1311 samples) from GH1, to 6831 training samples (tested on 359 samples) from GH2 (2015) and to 5431 training samples (tested on 285 samples) from GH2 (2016), and finally to 1096 training samples (tested on 57 samples) from GH3 (2018). 
Test data are not included in the training sets\footnote{The number of training samples available for models M1 and M2 vary as a result of extracting windows of one/two days from the data, scanning step size over the original dataset and the random extraction of test data, whose size is a fraction of the windowed dataset. The extraction algorithm can be found in the \textit{loaddataset.py} script in the github repository.}.

Model updates are performed on batches of 32 subsequent samples. 
As discussed above, in order to reduce catastrophic forgetting issues, an episodic memory system is employed, which replays samples together with the current batch when updating the model's weights.
Samples observed over time are stored into an episodic memory and retained following a prediction-error driven consolidation scheme.
In particular, a mechanism that chooses which samples to maintain in the episodic memory based on their expected contribution to the learning progress is employed. Each memory element consists of an input-output mapping, i.e. a fixed-length time series (of one day, or 288 samples, for model M1, or of two days, or 576, for model M2) of 6D vectors as input and a 2D vector (transpiration and photosynthesis) as output.
A memory element is also characterised by a prediction error 
-- i.e. how the model’s guess about this stored experience deviates from the actual measured value
-- and by an expected learning progress -- i.e. estimated as the absolute value of derivative of two subsequent prediction errors.
In particular, the learning progress LP is calculated as:

\begin{center}\begin{tabular}{ccl}\
$LP$ & $=$ & $|\epsilon_{t} - \epsilon_{t-1}|$\\
 & $=$ & $|(s^{*}_{t} - s_{t}) - (s^{*}_{t-1} - s_{t-1})|$
\end{tabular}\end{center}

where $\epsilon$ is the prediction error calculated at time $t$ or $t-1$ as the Euclidean distance between the sensory state $s$ (transpiration and photosynthesis) and the sensory prediction $s^{*}$. Sensory predictions (transpiration and photosynthesis) are inferred by feeding the 6D input of a memory element into the model.  
After each training iteration, all the memory elements are re-iterated and the associated derivative of the prediction error is updated.

We compare different architectures using three memory consolidation strategies and a \textit{no-memory} strategy - i.e. an architecture where no episodic memory system is employed - for each model, M1 and M2. 
The first strategy, hereon named \textit{discard high LP}, tends to consolidate memory elements that produced little variations in the prediction error. This is performed by discarding, at every memory update, the element characterised by the highest absolute value of the derivative of the prediction error (an estimate of the expected contribution to the learning progress) and by replacing it with the most recently observed sample. 
A second strategy, hereon named \textit{discard low LP}, tends to consolidate memory elements that produced big variations in the prediction error, likely to impact more on the learning progress during the next training iteration. In particular, it  discards the memory element characterised by the smallest variation in the prediction error. This strategy is more in line with the literature reviewed at the beginning of this paper, and we expect it to outperform the other strategy.
A third baseline strategy, hereon named \textit{discard random}, implements the standard memory consolidation approach in machine learning. In particular, at every memory update, it discards a randomly chosen sample from the memory.

Finally, we compare the different architectures varying another hyper-parameter, i.e. the probability of updating the memory: in a \textit{stable} configuration, the memory is updated every 5\% of the times a new sample is observed observations; in a \textit{plastic} configuration, this probability is set to 40\%, that is, the memory is updated much more frequently than in the \textit{stable} setup.
Table \ref{tab:doe} summarises the experiments carried out and their configurations.

\vspace{-0.5cm}
\begin{table}[h!]
\begin{scriptsize}
\begin{center}
\begin{tabular}{ |p{0.4cm}|p{0.9cm}|p{2.8cm}|p{2cm}|  }
 \hline
 \multicolumn{4}{|c|}{Design of experiments} \\
 \hline
  \hline
 ID & Model  & Mem. consolidation & Update prob. \\
 \hline
  \hline
 1  & M1 & No memory & - \\
 2  & M1 & Discard high LP & 0.05\% \\
 3  & M1 & Discard low LP & 0.05\% \\
 4  & M1 & Discard random & 0.05\% \\
 5  & M1 & No memory & - \\
 6  & M1 & Discard high LP & 0.4\% \\
 7  & M1 & Discard low LP & 0.4\% \\
 8  & M1 & Discard random & 0.4\% \\
 9  & M2 & No memory & - \\
 10  & M2 & Discard high LP & 0.05\% \\
 11  & M2 & Discard low LP & 0.05\% \\
 12  & M2 & Discard random & 0.05\% \\
 13  & M2 & No memory & - \\
 14  & M2 & Discard high LP & 0.4\% \\
 15  & M2 & Discard low LP & 0.4\% \\
 16  & M2 & Discard random & 0.4\% \\
 \hline
\end{tabular}
\vspace{-0.2cm}
\caption{The configurations of the experiments carried out in this work. Each experiment is run 10 times.}
\label{tab:doe}
\end{center}
\end{scriptsize}
\end{table}

\begin{figure*}
\captionsetup{font=footnotesize,labelfont=footnotesize}
\begin{minipage}[b][16.5cm]{0.24\linewidth}
\includegraphics[scale=0.23]{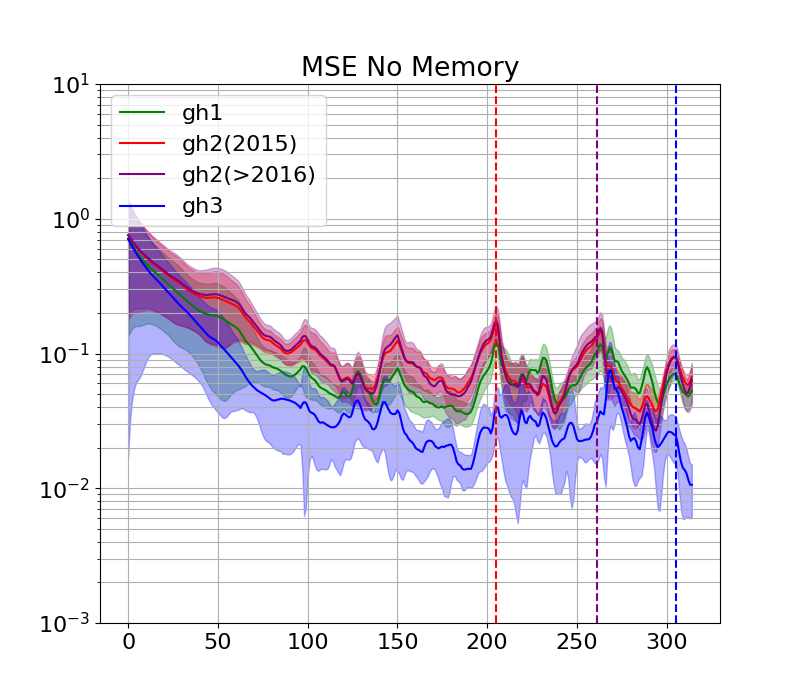}
\includegraphics[scale=0.23]{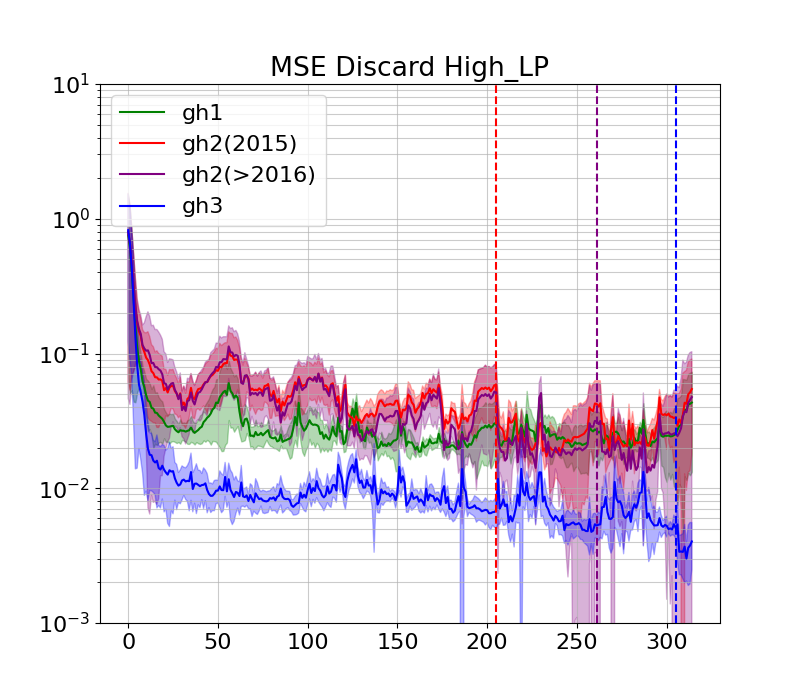}
\includegraphics[scale=0.23]{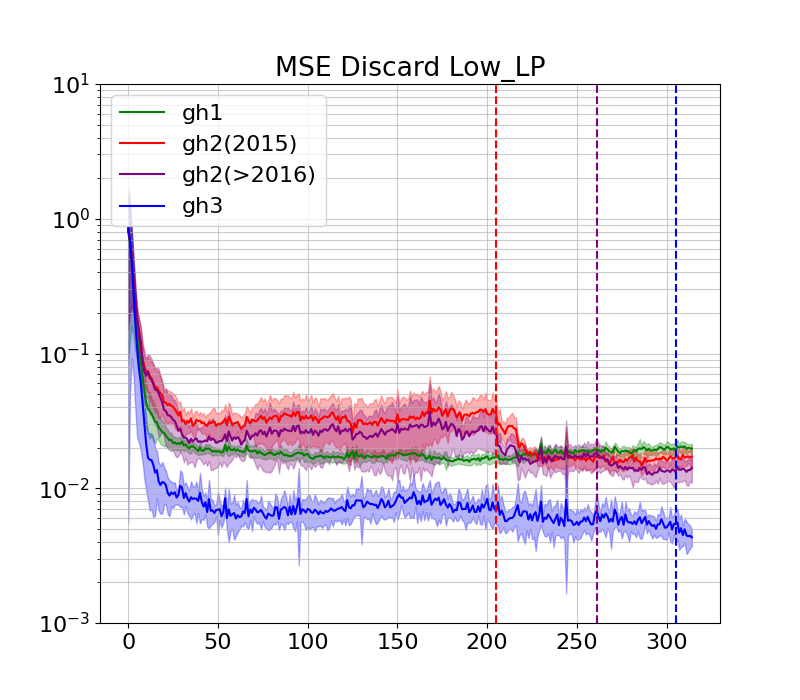}
\includegraphics[scale=0.23]{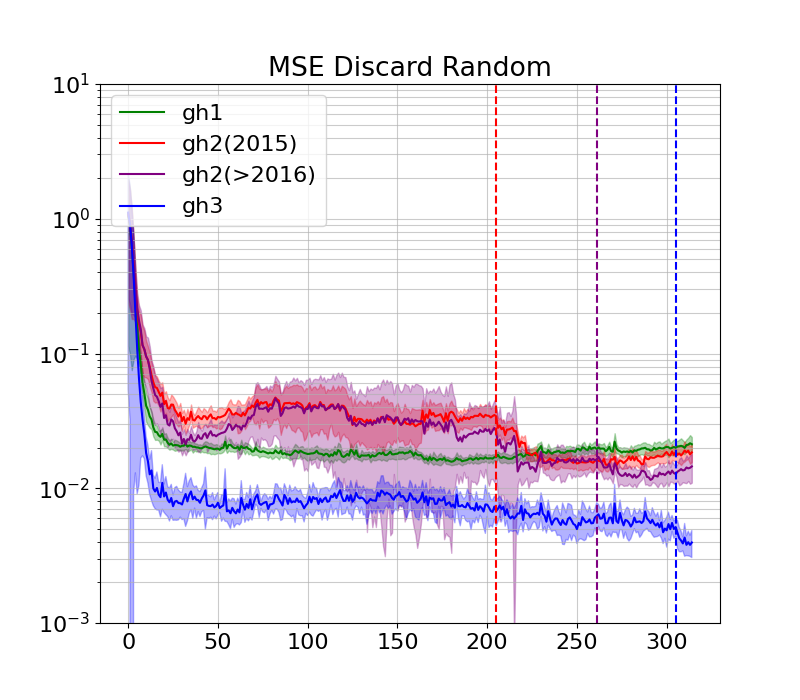}
\end{minipage}
\begin{minipage}[b][16.5cm]{0.24\linewidth}
\includegraphics[scale=0.23]{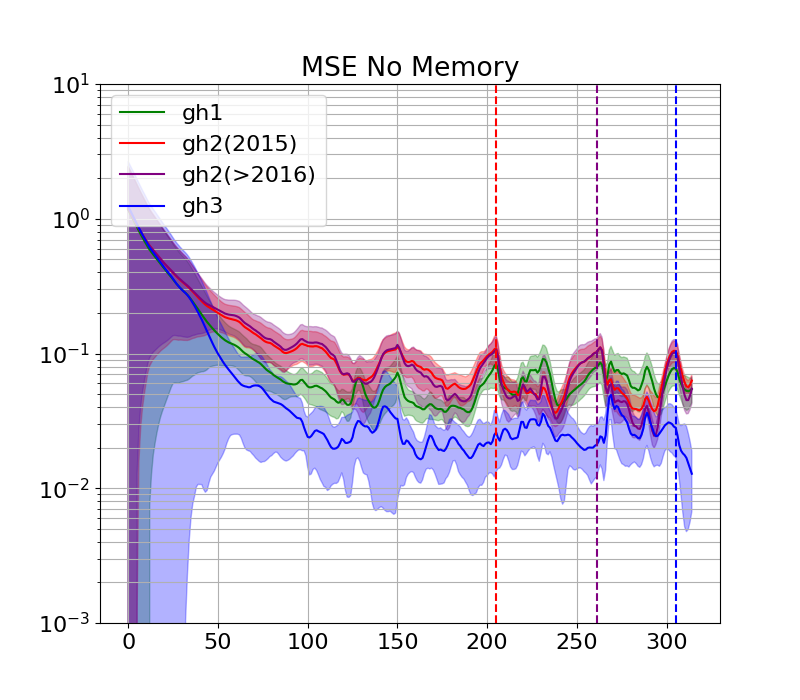}
\includegraphics[scale=0.23]{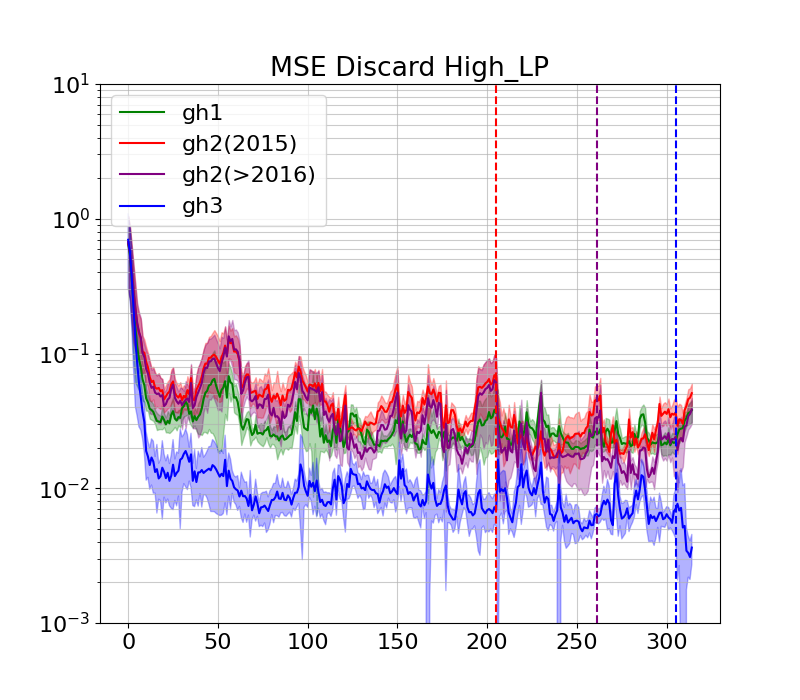}
\includegraphics[scale=0.23]{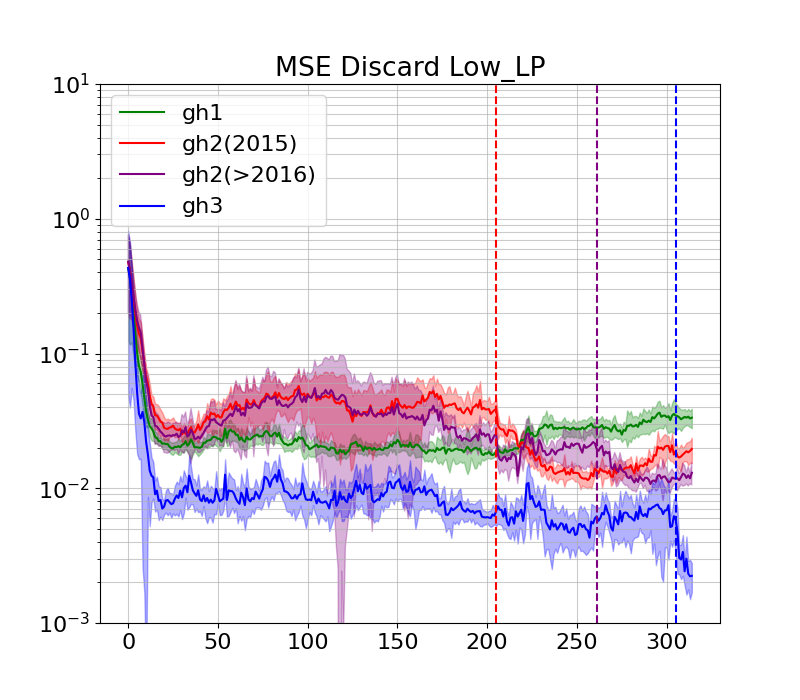}
\includegraphics[scale=0.23]{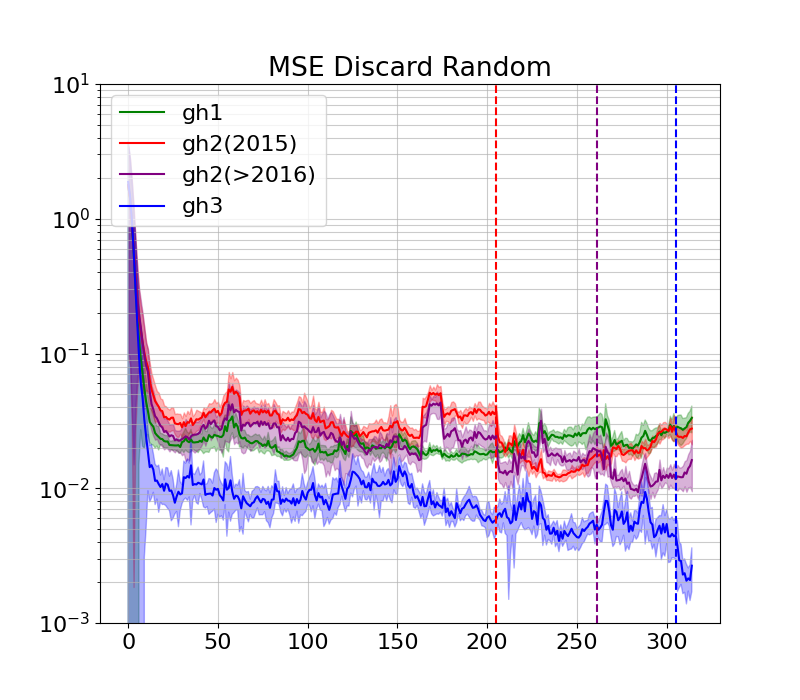}
\end{minipage}
\begin{minipage}[b][16.5cm]{0.24\linewidth}
\includegraphics[scale=0.23]{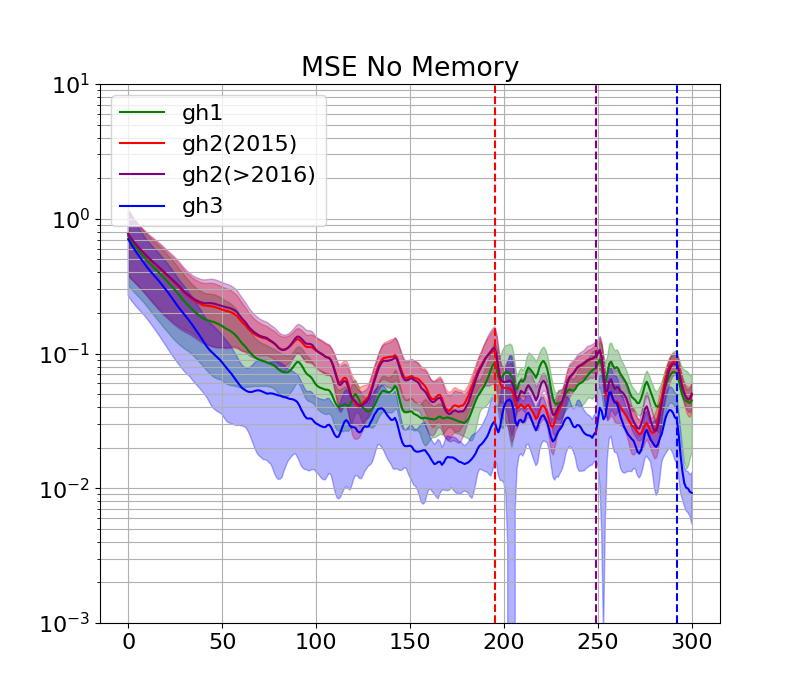}
\includegraphics[scale=0.23]{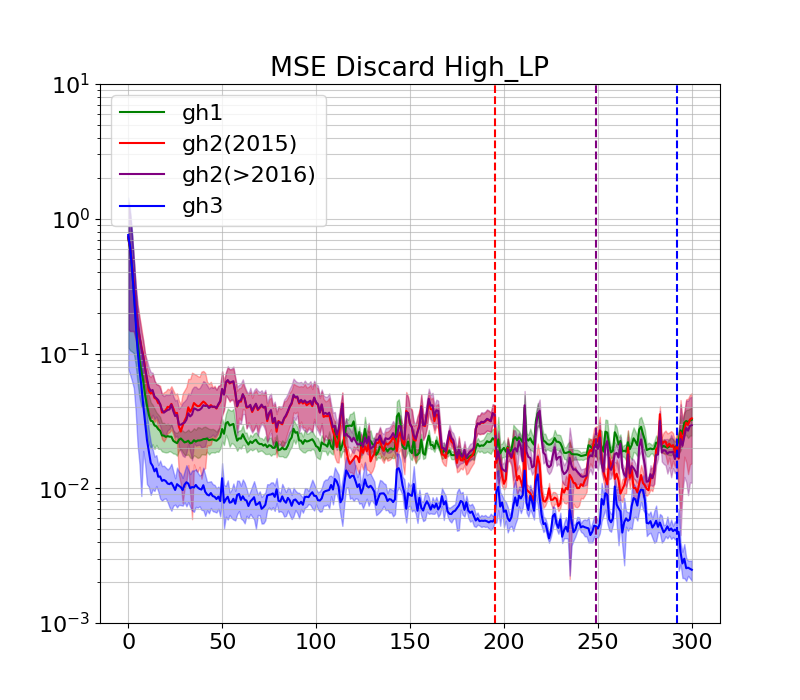}
\includegraphics[scale=0.23]{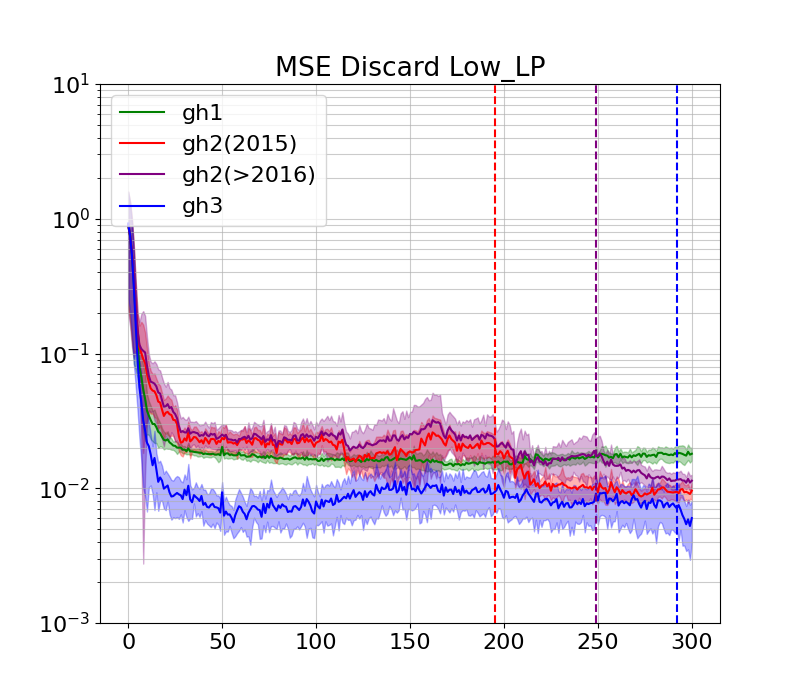}
\includegraphics[scale=0.23]{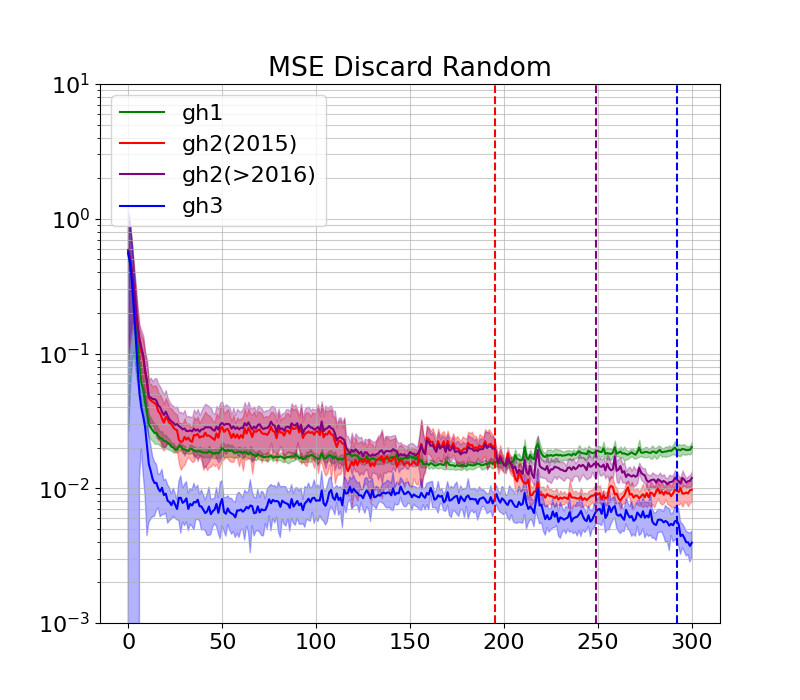}
\end{minipage}
\begin{minipage}[b][16.5cm]{0.24\linewidth}
\includegraphics[scale=0.23]{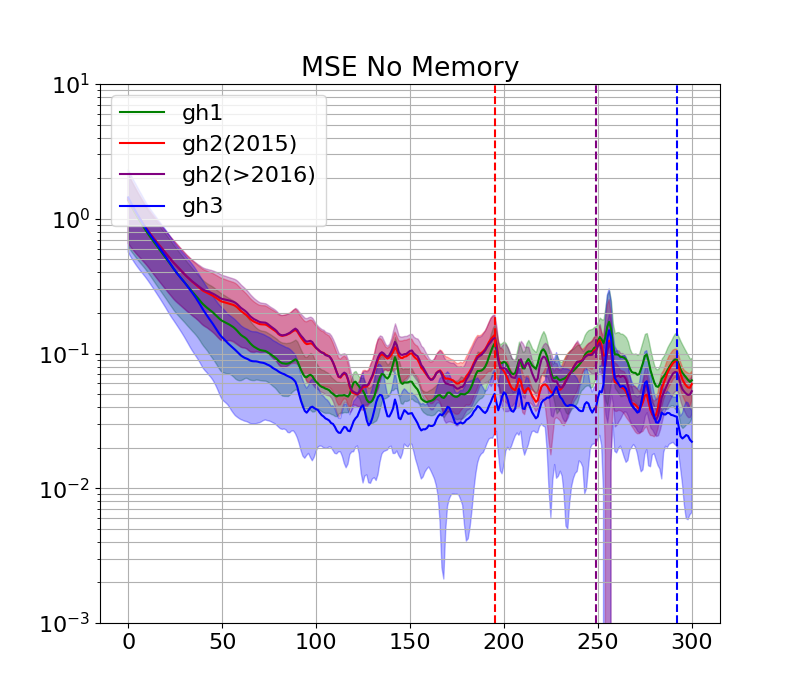}
\includegraphics[scale=0.23]{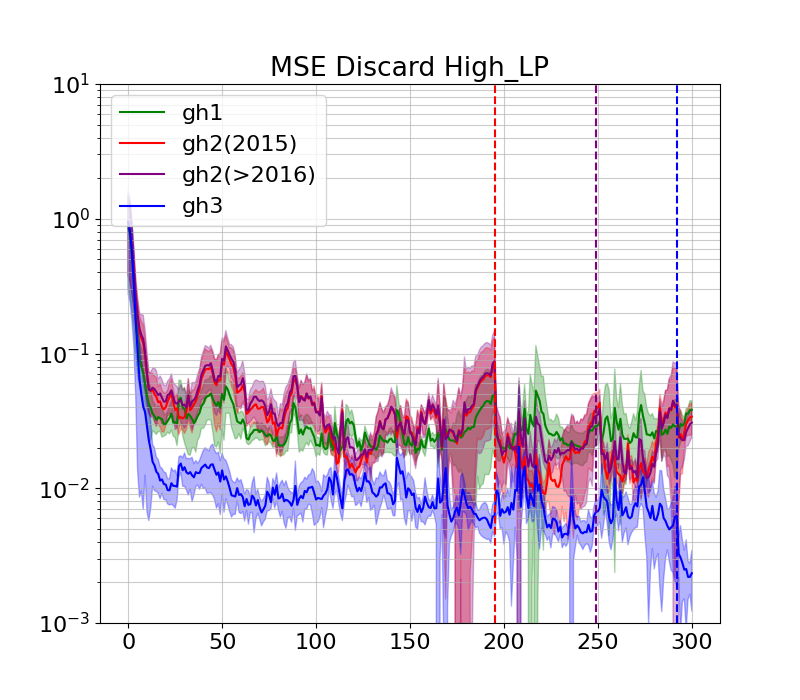}
\includegraphics[scale=0.23]{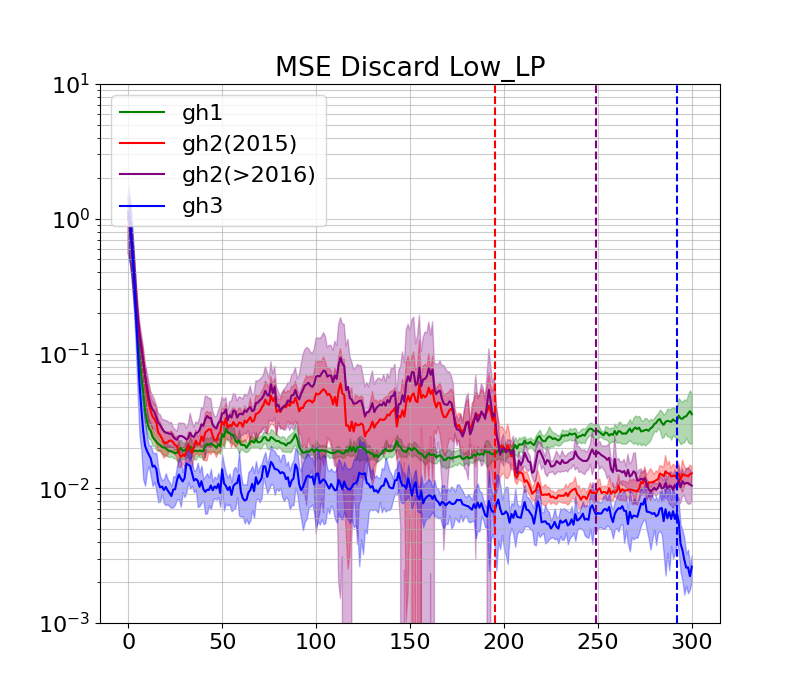}
\includegraphics[scale=0.23]{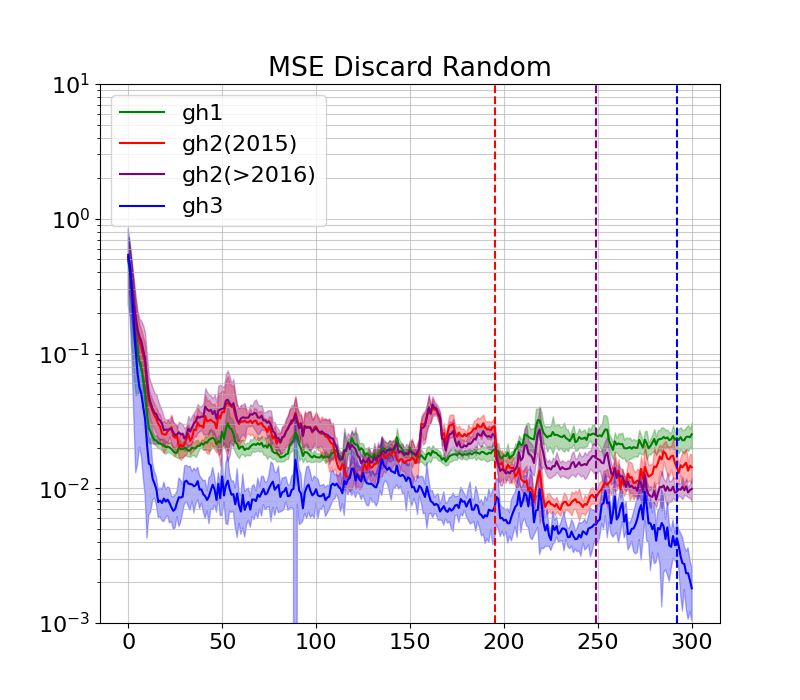}
\end{minipage}
\caption{Mean squared error over time of the following experiments: model M1 (input: one day of observations) in the \textit{stable} memory update configuration (first column, experiments 1-4 from top to down); model M1 in the \textit{plastic} memory update configuration (second column, exp.5-8 from top to down); model M2 (input: two days of observations) in the \textit{stable} memory update configuration (third column, exp.9-12 from top to down); model M2 in the \textit{plastic} memory update configuration (fourth column, exp.13-16 from top to down). First row shows the MSE of learning architecture when no episodic memory is employed; second and third rows show the MSE of the models employing the \textit{discard high LP} and \textit{discard low LP} memory consolidation strategies, respectively; the fourth row shows the MSE of the model employing the baseline \textit{discard random} consolidation strategy. Vertical axes indicate MSE values in the logarithmic scale. Horizontal axes represent time, in particular the iteration in which MSE has been estimated. Model update is performed every time a 32-batch of samples is observed. MSE is not computed at every model update, but rather with a slower pace, i.e. every four model updates. Vertical dashed lines indicate switches between training datasets. From time 0 to the iteration marked with the red vertical dashed line, the model is exposed to data recorded from GH1. From the iteration marked with the red line to the purple one, the model is exposed to data from GH2 (2015). From the iteration marked with the purple line to the blue one, the model is exposed to data recorded from GH2 (2016). Finally, from the instant marked with the blue line until the end, the model is exposed to data recorded from the production greenhouse GH3 (2018). 
Each experiment is repeated 10 times. Solid lines show the average MSE over the 10 runs calculated on four different test datasets (green plot: data from GH1, red plot: data from GH2, year 2015; purple plot: data from GH2, year 2016; blue plot: data from GH3). Shaded areas indicate errors (mean $\pm$ std.dev.) over the 10 runs.}
\label{fig:mse}
\end{figure*}

\section{Results}

Figure \ref{fig:mse} shows the mean squared error over time for each of the experiments depicted in Table \ref{tab:doe}. 
The absence of an episodic memory system (experiments 1, 5, 9, 13) produces big fluctuations in the MSE curves, likely due to catastrophic forgetting issues, and higher MSE values.
In these experiments, a sudden worsening in the model performance can be observed when training datasets are switched (see the peaks in the MSE near the vertical lines, exp. 1, 5, 9, 13), showing the poor adaptive capabilities of the model under this configuration\footnote{As noticed by one of the reviewers, spikes of the MSE in the \textit{no-memory} strategy already occur before a switch to a new dataset is performed. We believe this is due to a seasonal effect. As described in section Methodology, there are truncations in the time series during winter times.}. 

\begin{figure}
\captionsetup{font=footnotesize,labelfont=footnotesize}
\begin{minipage}[b][12.5cm]{0.49\linewidth}
\includegraphics[scale=0.23]{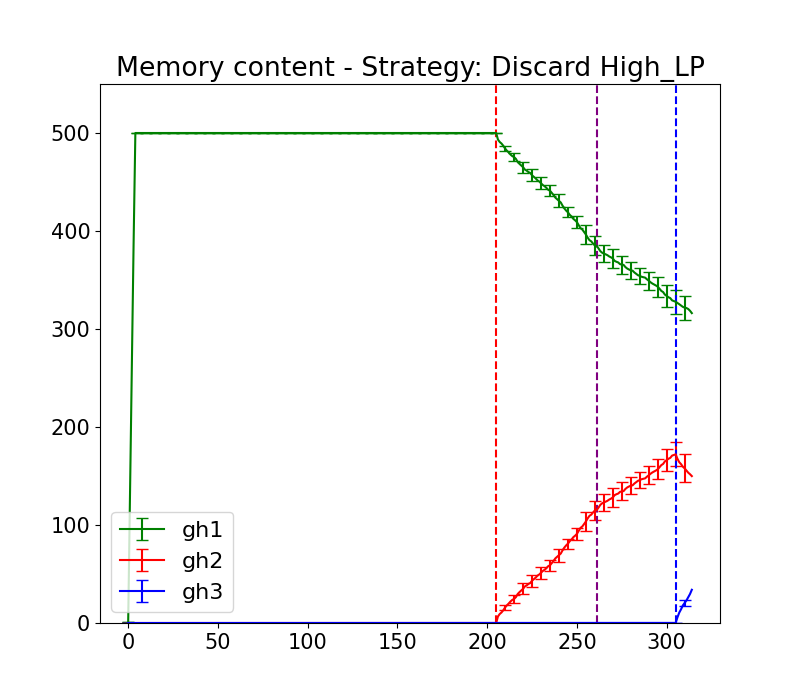}
\includegraphics[scale=0.23]{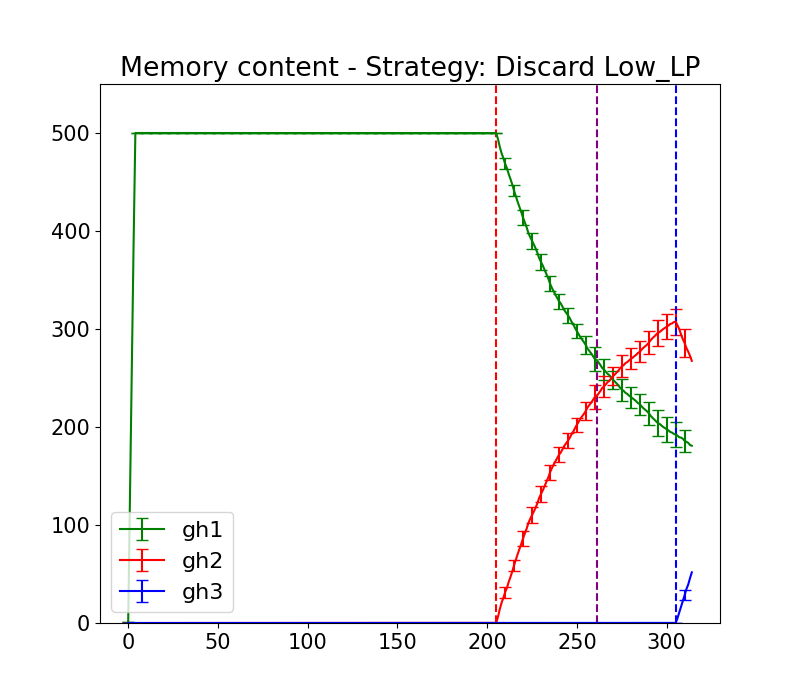}
\includegraphics[scale=0.23]{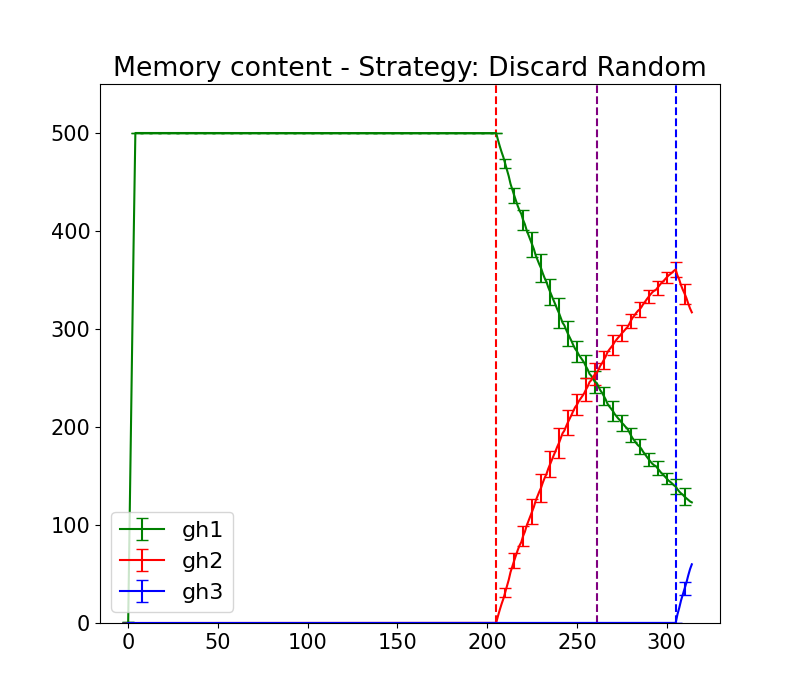}
\end{minipage}
\begin{minipage}[b][12.5cm]{0.45\linewidth}
\includegraphics[scale=0.23]{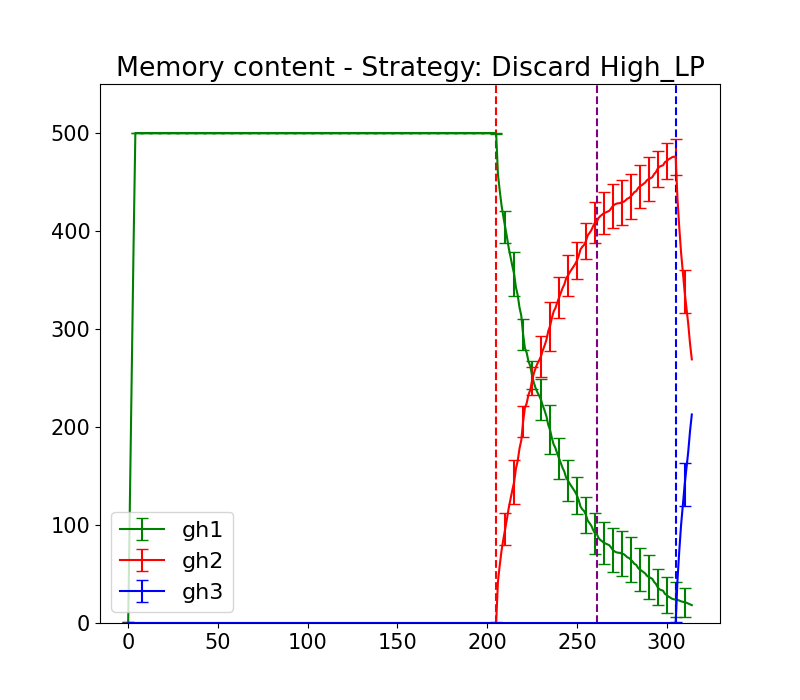}
\includegraphics[scale=0.23]{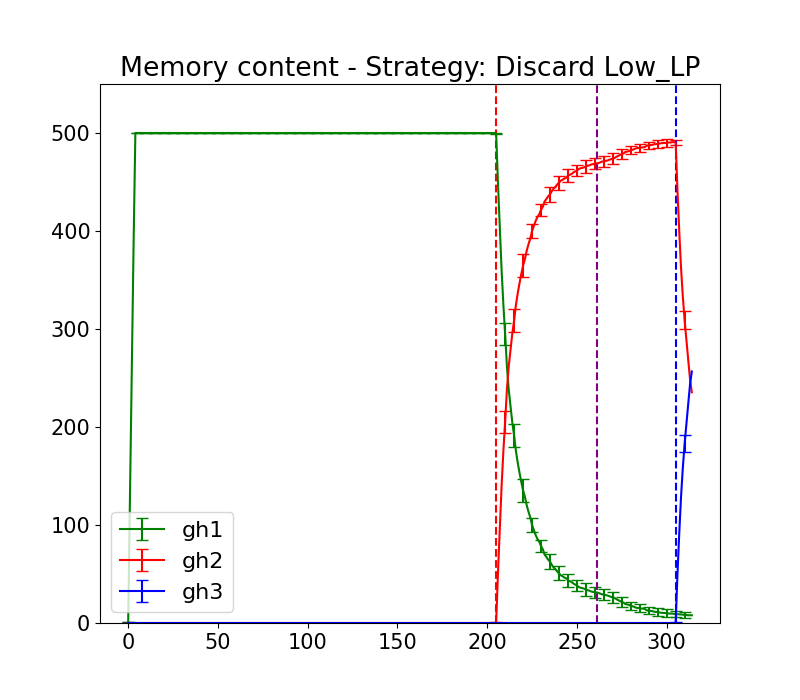}
\includegraphics[scale=0.23]{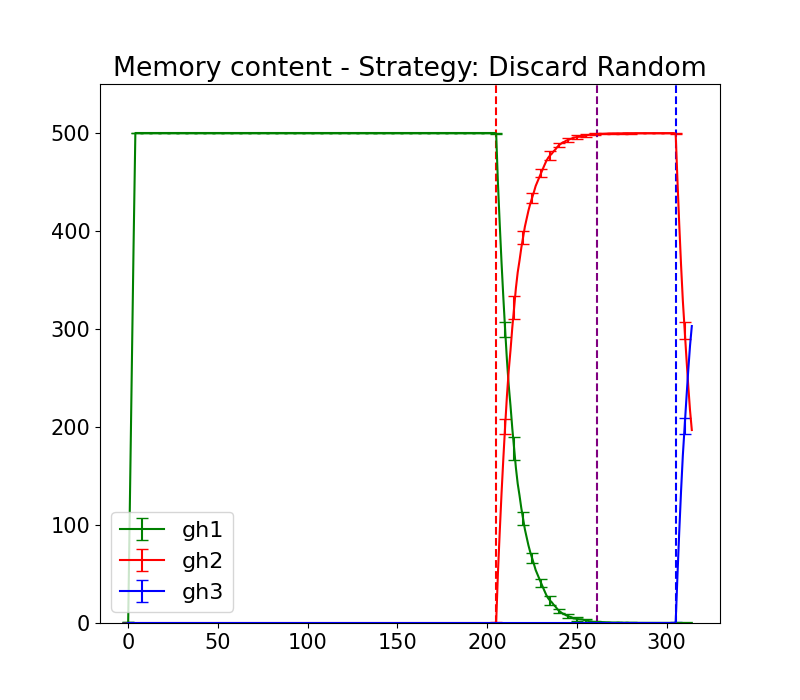}
\end{minipage}
\caption{Content of the episodic memory over time (horizontal axis) for each experiment adopting memory consolidation strategies (means $\pm$ std.dev. over 10 runs per experiment). y-values represent the amount of elements (from 0 to 500, where 500 is memory size) for each dataset in the memory (GH1: green, GH2: red, GH3: blue). Horizontal axes represent time. Plots on the left refer to \textit{stable} configurations (exp.2-4 from top to down). Plot on the right to \textit{plastic configurations} (exp.6-8, from top to down).
Experiments 10-12 and 14-16 have been omitted, as plots closely resemble those shown here.}
\label{fig:mem_content}
\end{figure}

On the contrary, employing an episodic memory system produces a more stable learning progress (see second, third and fourth rows in Figures \ref{fig:mse}). As expected, the \textit{discard low LP} memory consolidation strategy outperforms the other methods. The model under this configuration shows more stability despite the changes in the training distributions. 
The \textit{discard high LP} strategy, in fact, seemingly over-consolidates past and, perhaps, not much informative  experiences (see later comment about the variance of the stored episodic memories). This can be noticed in Figures \ref{fig:mem_content}, which illustrate the content of the episodic memory over time for the \textit{stable} (left column) and \textit{plastic} (rigth column) configurations. In particular, the plots show how many elements from GH1 (green), GH2 (red) and GH3 (blue) have been stored in the memory over time\footnote{The maximum size of the memory is set to 500 elements in all the experiments (vertical axis in the plots). Memory is filled up with any observed sample, until it is full. Thereafter, the chosen consolidation strategy is applied. We tested different memory sizes (in the range 100-1000, empirically chosen). Clearly, this slightly impacted performances as depicted in Figures \ref{fig:mse} and \ref{fig:mem_content}, although not affecting the final conclusions of this paper. We did not report the studies here, to not increase considerably the number of experiments presented.
We decided to fix the memory size to 500, as to allow showing clear differences between consolidation strategies.}. As evident from the figures, the \textit{discard low LP} strategy fills the memory with new samples faster than the \textit{discard high LP} strategy. It present a similar trend as the \textit{discard random} strategy. Nonetheless, \textit{discard low LP} strategy maintains more memory elements from past greenhouses than \textit{discard random} strategy (see, for instance) the red curves in Figure \ref{fig:mem_content}, between exp. 7 and 8: after the purple vertical line (model exposed to GH2 2016 training data), there is almost no sample from GH1 left in the memory using the \textit{discard random} strategy. This impacts also models' performance, as it can be seen from Figure \ref{fig:mse}. For instance, it can be noted in experiment 15 and 16, that performance on test dataset from GH2 (2015, red curve), after exposing the models to GH2(2016) data (vertical purple dashed line), degrades faster in the \textit{discard random} strategy than in the \textit{discard low LP} one. Similar effect can be observed between experiments 7 and 8. 

\begin{figure}
\captionsetup{font=footnotesize,labelfont=footnotesize}
\begin{minipage}[b][8.5cm]{0.5\linewidth}
\includegraphics[scale=0.23]{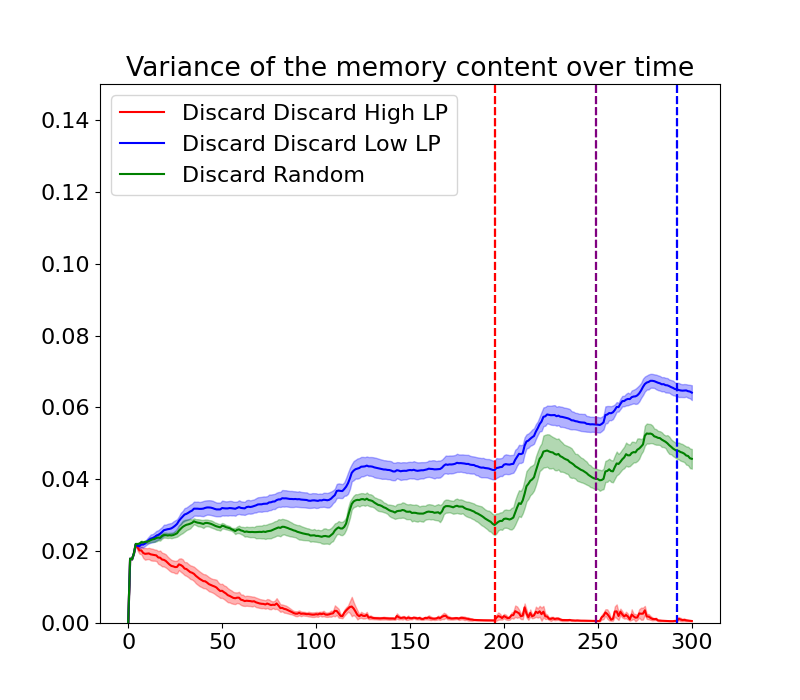}
\includegraphics[scale=0.23]{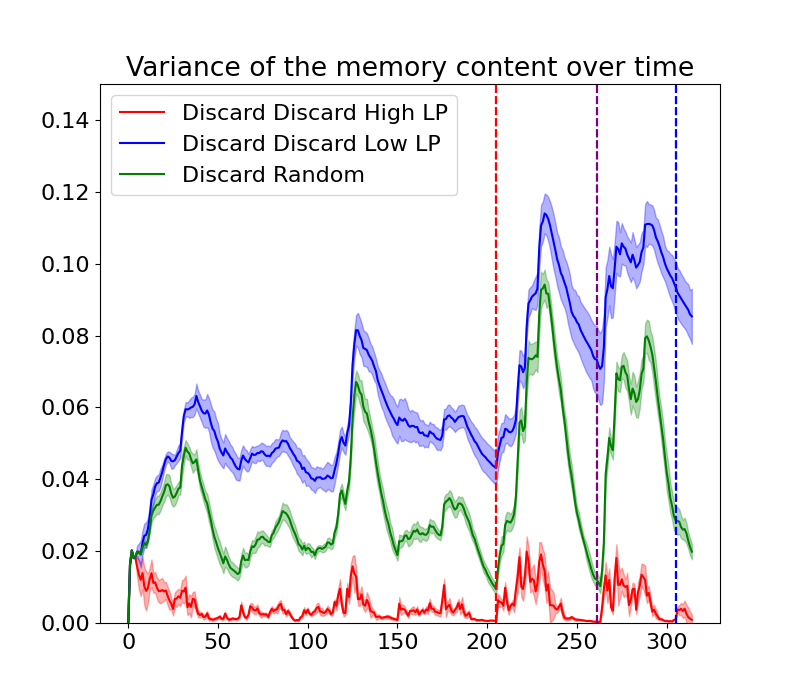}
\end{minipage}
\begin{minipage}[b][8.5cm]{0.48\linewidth}
\includegraphics[scale=0.23]{m2_stable_mem_proportions.png}
\includegraphics[scale=0.23]{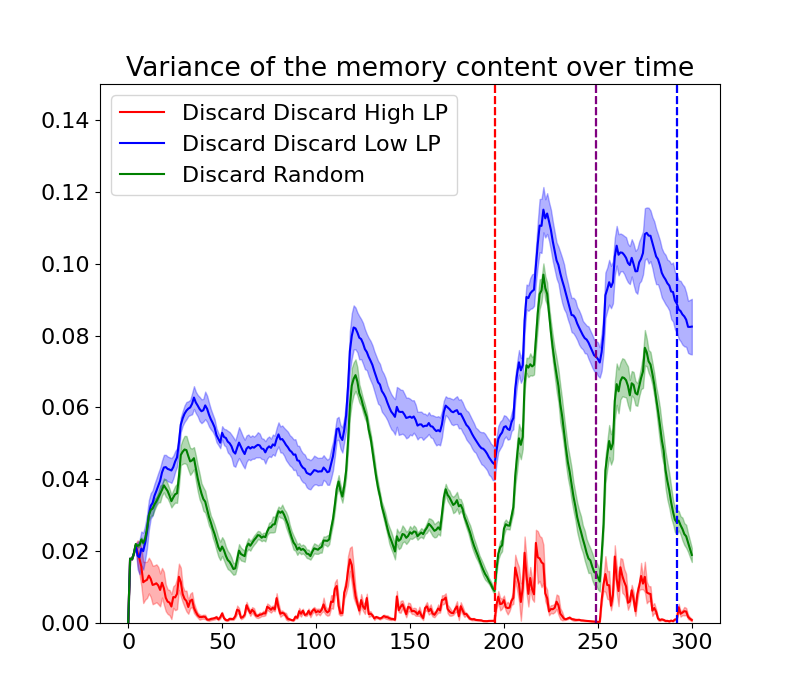}
\end{minipage}
\caption{The variance of the content of the memory over time (considering only the output features, i.e. transpiration and photosynthesis values). Left column (top to down): M1-stable and M1-plastic; second column: M2-stable and M2-plastic}
\label{fig:output_variance}
\end{figure}

Replaying more recent samples during the model update is likely to increase the plasticity of the system. In fact, smaller peaks in the MSE in the \textit{discard low LP} plots can be observed when the distribution changes. \textit{Plastic} configurations in general respond faster to new data (see for instance the blue lines at the end of experiments 7, 8, 15 and 16).
Moreover, the \textit{discard low LP} strategy ensures that a higher variance in the values stored in the memory is maintained over time, as compared to the \textit{discard high LP} and \textit{discard random }strategies. This can be seen from Figures \ref{fig:output_variance}. We believe that this is providing a good balance between stability and plasticity of the model.

\begin{figure}
\vspace{-0.4cm}
	\centering
	\includegraphics[width=5cm]{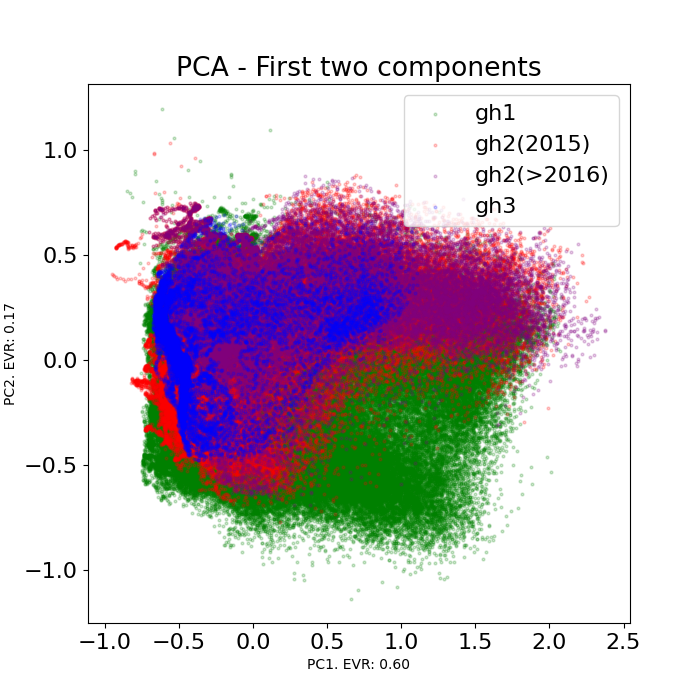}
\vspace{-0.4cm}
	\caption{Principal component analysis of all datasets.}
	\label{fig:pca}
\end{figure}

Finally, it is important to highlight that a principal component analysis (PCA) 
carried out on all the datasets (Figure \ref{fig:pca}), and estimated on 8 dimensions -- i.e. six sensors data, and transpiration and photosynthesis --  shows a partial overlap between datasets. GH2 and GH3 data seem to be partly represented on GH1 data. As evident from Figure \ref{fig:mse}, the MSE curves for GH3 test data (blue curves) during the first learning phase -- i.e. where data from GH1 are used -- show a steeper descent compared to the others, although no data from GH3 is being learned by the network, yet. This can be explained by greenhouses GH1 and GH2 being of the same construction and location, while GH3 is bigger and subject to different meteorological conditions. 
Additionally, GH1 was used to test a number of climate control strategies, resulting in a broader range of conditions being reflected in the data.
We believe that more heterogeneous datasets would have emphasised the advantages of using the proposed approach in terms of adaptivity. Nonetheless, despite the similarity of the datasets, the proposed memory consolidation strategies clearly demonstrate to produce stable learning systems, more than standard memory consolidation strategies.

\section{Conclusions}
This paper presented an architecture in which episodic memory replay and prediction-error driven  consolidation are used to tackle online learning in deep recurrent neural networks. Inspired by evidences in cognitive sciences and neuroscience, memories are retained depending on their congruence with the prior knowledge stored in the system.
 This congruence is estimated in terms of prediction error resulting from a generative model, a deep recurrent neural network. This approach produces a good balance between stability and plasticity in the model.

Importantly, this work aimed also at transferring this AI strategy onto an application for the greenhouse industry, i.e. the transfer of climate models from research facilities to production greenhouses. This technical possibility can greatly increase the value of research-generated data, which could then be supplied in addition to neural models for direct use in the industry. We show that the system exposed to data recorded from a research greenhouse can be transferred to a production facility, without facing the need to re-train on a big amount of data from the new setup, a process that is costly and involves a high risk of damaging the crop.

This demonstrates that some of the paradigms of developmental robotics and of brain-inspired computational modelling can start to be transferred from laboratories to innovative applications in the industry.


\bibliographystyle{spmpsci}      
\bibliography{biblio}   

\end{document}